\documentclass[sigconf,nonacm]{acmart}

\usepackage{algorithmic}
\usepackage{graphicx}
\usepackage{textcomp}
\usepackage{xcolor}
\usepackage{xspace}
\usepackage{soul}
\usepackage{url}
\usepackage{multirow}
\usepackage{hyperref}
\usepackage{tcolorbox}
\usepackage{float}
\usepackage{booktabs}
\usetikzlibrary{positioning, arrows.meta, shapes.geometric}

\usepackage{tikz}
\usetikzlibrary{positioning, arrows.meta, shapes}
\usepackage{fontawesome5}
\usepackage{enumitem}

\definecolor{cardblue}{HTML}{2B6CB0}
\definecolor{cardbg}{HTML}{EBF4FA}
\definecolor{cardyellow}{HTML}{FFFBE6}
\definecolor{yellowborder}{HTML}{D4A017}

\def\BibTeX{{\rm B\kern-.05em{\sc i\kern-.025em b}\kern-.08em
    T\kern-.1667em\lower.7ex\hbox{E}\kern-.125emX}}

\newcommand{\rqone}{\textit{What do developers show in the visual artifacts of their PR descriptions?}}
\newcommand{\rqtwo}{\textit{Why do developers include visual artifacts in their PR descriptions?}}
\newcommand{\rqthree}{\textit{How does code review differ for PRs whose descriptions include visual artifacts?}}

\usepackage{xspace}
\newcommand{\dataset}{\textit{VisPR}\xspace}

\newif\ifcomments
 \commentstrue       

\ifcomments
    \newcommand{\yuan}[1]{\textcolor{red}{[Yuan: #1]}}
    \newcommand{\AM}[1]{\textcolor{blue}{[amit: #1]}}
    \newcommand{\MH}[1]{\textcolor{orange}{[Max: #1]}}
\else
    \newcommand{\yuan}[1]{}
    \newcommand{\AM}[1]{}
    \newcommand{\MH}[1]{}
\fi

\newcommand{\ct}[1]{\textit{#1}}   
\newcommand{\pp}[1]{\textit{#1}}   

\begin{document}

\title{An Empirical Study of Visual Artifacts in Pull Request Descriptions}

\begin{abstract}
Reviewers rely on pull request (PR) descriptions to understand what a change does, yet some changes, such as UI updates or runtime behavior, are hard to convey in text alone. PR authors attach visual artifacts to fill this gap, but how often they do so, what they show, and whether they change how a PR is reviewed remain unknown. We present a large-scale study of visual artifacts in GitHub PR descriptions. From 6,505,113 closed PRs in 5,012 repositories, we built \dataset, a dataset of 2,473,892 English, human-authored PRs from 647 repositories where visual artifacts are an established practice. Of these, 329,185 (13.3\%) are visual PRs, i.e., their descriptions contain at least one visual artifact (479,974 in total). By manually labeling a sample of 1,113 visual artifacts from 891 merged visual PRs, we derived two taxonomies: 11 content types capturing \emph{what} developers show and six purposes capturing \emph{why} they include visual artifacts. The dominant content type is  \emph{Interaction \& State}, which shows how the interface responds to user actions. its share rises from 37.0\% of images to 63.5\% of GIFs and 79.7\% of videos, whereas about a quarter of images (23.4\%) show text such as logs, code, or error messages . Compared with 1,782 merged non-visual PRs matched on repository, PR type, and change size, visual PRs receive 43\% more review comments and 17\% more reviewers, and take 72\% longer from first response to merge, time to merge and churn-normalized review efficiency do not differ significantly. LLM-assisted classification of 10,031 review comments in two groups shows that review centers on code in both, but comments on visual PRs address how the UI looks or behaves twice as often (6.4\% vs.\ 3.3\%), and visual-output fixes after review are about four times as common (4.7\% vs.\ 1.2\% of PRs). Visual artifacts are thus associated with closer scrutiny of what users see, not with an overall faster review. Our dataset, taxonomies, and findings offer guidance for PR authors and maintainers, hosting platforms, AI coding agents, and builders of multi-modal coding benchmarks.

\end{abstract}


\author{Tianzi Hao}
\affiliation{%
  \institution{Queen's University}
  \city{Kingston}
  \country{Canada}}
\email{20th32@queensu.ca}

\author{Kazi Amit Hasan}
\affiliation{%
  \institution{Queen's University}
  \city{Kingston}
  \country{Canada}}
\email{kaziamit.hasan@queensu.ca}

\author{Yuan Tian}
\affiliation{%
  \institution{Queen's University}
  \city{Kingston}
  \country{Canada}}
\email{y.tian@queensu.ca}

\author{Bram Adams}
\affiliation{%
  \institution{Queen's University}
  \city{Kingston}
  \country{Canada}}
\email{bram.adams@queensu.ca}
\maketitle

\section{Introduction}\label{intro}

Pull request (PR) descriptions are the first point of communication between the PR author and reviewers about a code change. A description explains what the change does and why it is needed, and helps reviewers understand it before they examine the code~\cite{liu2019automatic, ram2018makes}. Prior studies examined PR descriptions from several perspectives. Ram et al.~\cite{ram2018makes} found that the change description is one of the factors that make a change easier to review. Pirouzkhah et al.~\cite{pirouzkhah2026value} analyzed 80K PRs and found that stating the type of feedback the author expects best predicts whether a PR is accepted and how much reviewers engage with it. Liu et al.~\cite{liu2019automatic} found that more than 34\% of over 333K PRs have an empty description and proposed an approach to generate descriptions automatically. More recent studies examined descriptions produced by AI tools. PRs whose descriptions were partly generated by GitHub Copilot required less review time and were more likely to be merged~\cite{xiao2024generative}, and the description styles of AI coding agents are associated with differences in reviewer engagement, response time, and merge outcomes~\cite{watanabe2026how}. These studies focus on the text of PR descriptions. However, a description can also contain images, GIFs, and videos, which GitHub lets authors embed by dragging, selecting, or pasting a file~\cite{github_attaching_files}.\footnote{For example, \url{https://github.com/apecloud/kubeblocks/pull/2132} and \url{https://github.com/blockprotocol/blockprotocol/pull/133}.} We call these \emph{visual artifacts}, and a PR whose description contains at least one is a \emph{visual PR}. Visual artifacts convey what is difficult to express in text. The text describes a change and the diff shows which lines changed, but a screenshot can show how the user interface looks after the change, and a GIF or video can show how the application behaves. Developers use them widely. Some projects request screenshots in PR templates,\footnote{\url{https://github.com/kodadot/nft-gallery/pull/4374}} and visual artifacts appear in PRs of 4,908 of the 5,012 repositories we collected (97.9\%, Table~\ref{tab:dataset}).

Despite this widespread use, visual artifacts in PRs have received little attention. Pirouzkhah et al.~\cite{pirouzkhah2026value} listed screenshots as a recommended element of PR descriptions but excluded them from their analysis because they apply only to some PRs, such as those that change the user interface. Research on visual artifacts has instead focused on issue reports. Kuramoto et al.~\cite{kuramoto2024understanding} studied issue reports in 34 GitHub repositories and found that those with images or videos are more likely to report bugs (44.6\% for images and 71.1\% for animations and videos, versus 32.2\%), receive their first comment sooner, and complete the conversation faster, and that their visuals mainly show program behavior and the user interface to introduce problems. Tan et al.~\cite{imageR2025} labeled 6,235 Bugzilla issue reports by the type of attached image and built a tool that predicts when an issue report needs an image (F1 = 0.76). Benchmarks for AI coding agents also take their visual evidence from issues. Images are necessary to solve 83.5\% of the tasks in SWE-bench Multimodal~\cite{swebench_multimodal}, and removing them reduces file-level localization accuracy by 4.4--4.9 points in MM-IssueLoc~\cite{zhan2026mmissueloc}. These findings do not transfer to PRs, because the visual artifact plays a different role. In an issue report, it shows a problem that has not yet been fixed. In a PR description, it accompanies a change that reviewers must evaluate and can show the result of that change, the state before it, or both. 


To address this gap, we present a large-scale empirical study of visual artifacts in GitHub PR descriptions. We collected 6,505,113 closed PRs from 5,012 repositories, retained those written in English, and labeled each with its author type (human, bot, or AI agent) and the visual artifacts in its description. We then selected 647 repositories in which visual artifacts are an established practice, each with at least 100 human-authored visual and 100 non-visual PRs. In these repositories, 95.7\% of the visual artifacts in bot PRs are SVG files such as status badges, and AI agents authored only 0.3\% of visual PRs. We therefore focus on the 2,473,892 human-authored PRs, 329,185 of them visual, which form \dataset.

We ask what developers show in these visual artifacts (RQ1), why they include them (RQ2), and how the review of visual PRs differs from similar non-visual PRs (RQ3). The answers can inform how projects and developers use visual artifacts, how platforms and AI agents produce and read them, and how benchmarks use them as evidence of a correct change. We manually labeled 1,113 visual artifacts across images, GIFs, and videos and compared the review of the 891 visual PRs containing them with 1,782 matched non-visual PRs, using an LLM to classify their 10,031 issue, review, and inline comments. What developers show depends strongly on the media format. The dominant content type, \textit{Interaction \& State}, which shows what happens during or after a user action, accounts for 37.0\% of images but 63.5\% of GIFs and 79.7\% of videos, whereas about a quarter of images (23.4\%) show text such as logs, code, or error messages. Most visual artifacts show new functionality that the PR adds (\textit{Introduce}) or that a fix or change works as intended (\textit{Verify}), and only 6.8\%--15.9\% show the state before the change. Visual PRs receive 43\% more comments and 17\% more reviewers and take 72\% longer from the first response to merge, yet the commonly used churn-based efficiency metric detects no difference. Review centers on code in both groups, but comments on visual PRs address how the UI looks or behaves twice as often (6.4\% vs.\ 3.3\%), and visual-output fixes after review are about four times as common (4.7\% vs.\ 1.2\% of PRs). Visual artifacts are thus associated with closer scrutiny of what users see, not with faster review. This paper makes the following contributions:
\begin{enumerate}
    \item A dataset of 2,473,892 English, human-authored PRs from 647 GitHub repositories, in which each PR is labeled with the visual artifacts in its description, if any.
    
    \item We propose two taxonomies of visual artifacts in PR descriptions, with 11 content types for what developers show and six purposes for why they show it.
    
    \item We compare the review of 891 visual PRs with 1,782 matched non-visual PRs and show that visual PRs receive more comments, reviewers, and scrutiny of what users see, with visual-specific comments twice as common and visual-output fixes four times as common, whereas time to merge and review efficiency do not differ significantly.
\end{enumerate}

\section{Dataset and Study Scope}\label{data}
\begin{figure*}[t]
    \centering
    \resizebox{\textwidth}{!}{%
        \input{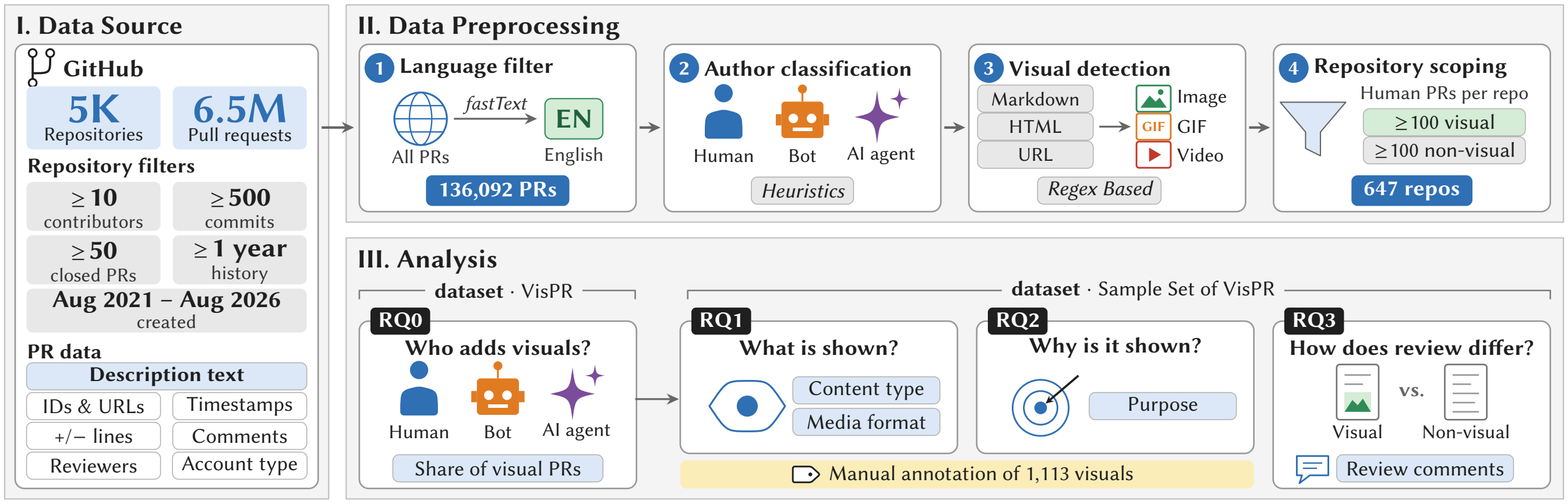}
    }
    \caption{Overview of our study design.}
    \label{fig:overview}
\end{figure*}

In this section, we first describe how we create \dataset for studying visual artifacts in GitHub PR descriptions through data collection (Section~\ref{sec:sourcedata}) and preprocessing (Section~\ref{sec:preprocessing}). We then present a preliminary analysis of who (humans, bots, or AI agents) embeds visual artifacts and in which formats (Section~\ref{sec:preliminary}), and explain how we derived the scoped dataset used to answer RQ1--RQ3 (Section~\ref{sec:sample_set}). Figure~\ref{fig:overview} shows an overview of our study, and Table~\ref{fig:overview} shows basic statistics for the dataset that we created.

\subsection{Data Collection}
\label{sec:sourcedata}
\textbf{Repository selection.} We identified candidate repositories using the SEART GitHub Search Engine~\cite{dabic2021sampling}. Following Ouf et al.~\cite{ouf2026dogood}, we selected repositories with at least 10 contributors, 500 commits, 50 closed PRs, and one year of development history, we further  restricted the selection to repositories created between August 2021 and August 2026, to focus on recent and actively developed projects. The search, conducted in August 2026, collected 5,012 repositories. 

\textbf{Closed PR collection.} We further collected PR data from the 5,012 repositories using the GitHub GraphQL API~\cite{github_graphql_api}. For each repository, we retrieved all closed PRs, whether merged or not, so that every PR in the dataset had reached its final state. We recorded three groups of data:
(1)~\emph{repository metadata}: number of stars, primary programming language, and license.
(2)~\emph{PR attributes}: identifiers (ID, number, and URL); creation, closing, and merge times, size (added lines, deleted lines, and changed files) and the description, which is the primary focus of our analysis, and (3)~\emph{comments and participants}: the text of issue comment, and the accounts of the PR author, the commenters, and the person who merged the PR, together with the type of each account (User or Bot). In total, we collected 6,505,113 PRs.

\subsection{Data Preprocessing}
\label{sec:preprocessing}

\textbf{Language filtering.} To ensure that all analyzed text is in the same language, we retained only PRs written in English. Following Kuramoto et al.~\cite{kuramoto2024understanding}, we detected the language of each PR by applying a pre-trained fastText model~\cite{joulin2016bag} to its title, description, and issue comments. This step removed 136,092 PRs (2.1\%), leaving 6,369,021 PRs. 

\textbf{Author type classification.} We classified the author of each PR as a human, a bot, or an AI agent, adapting the heuristics of Romero-Arjona et al.~\cite{romeroarjona2026whenai}. An account is labeled as an AI agent if its username exactly matches, or contains, a known agent account name from the lists published by them. Otherwise, it is labeled as a bot if GitHub reports its account type as Bot in Section~\ref{sec:sourcedata} or its username contains the suffix[bot], and as a human in all other cases, Because AI agents can also open PRs from human or generic bot accounts, we then checked two PR-level signals of agent authorship, either of which overrides the account label. First, we matched the PR source branch against the default branch prefixes of AI agents(e.g.,codex/, copilot). Second, we searched the PR title and description for attribution phrase that combine an authorship verb (e.g., "generated" or "implemented"), a preposition (e.g.,"by","with"), and an agent name. 

\textbf{Visual artifact detection.} Following Kuramoto et al.~\cite{kuramoto2024understanding}, we define a \emph{visual artifact} as an image, animated GIF, or video embedded in a PR description. We extend their detection in two ways: we cover more file formats within each media type, and we recognize more ways in which media are represented in a description. To determine these representations, we manually tested every method GitHub provides for attaching media (dragging and dropping, selecting a file, and pasting from the clipboard) with every image, GIF, and video format GitHub supports~\cite{github_attaching_files}. Media appear in three forms: Markdown image syntax, HTML tags (e.g., \texttt{<img>} and \texttt{<video>}), and plain URLs. The first two forms can be identified by their syntax, a plain URL represents media only when it points to a media file, which must contain file extensions in its url to be collected (e.g., \texttt{.png} or \texttt{.mp4}) or by its GitHub upload path. We implemented one regular expression per form and applied all three to every PR description. We restrict detection to descriptions because the description is the author's account of the change, presented to reviewers before review begins.

\textbf{Creating \dataset.} \label{sec:scoping}
Visual artifacts appear in PRs of 4,908 of the 5,012 repositories (97.9\%). The presence of a visual artifact, however, does not show that a repository uses visual artifacts as a regular practice. We therefore retained repositories with at least 100 human-authored visual PRs and at least 100 human-authored non-visual PRs. The first criterion ensures that visual artifacts are an established practice rather than incidental and the second ensures that each repository supplies enough non-visual PRs to serve as controls in RQ3. This yielded 647 repositories containing 2,473,892 human-authored PRs, which form \dataset. Of these, 329,185 (13.3\%) are visual PRs.

\begin{table}[t]
\centering
\caption{Dataset after each construction step.}
\label{tab:dataset}
\begin{tabular}{lrr}
\toprule
Step & Repositories & Pull requests \\
\midrule
Initial collected dataset            & 5,012 & 6,505,113 \\
English-language PRs              & 5,012 & 6,369,021 \\
\quad with any visual artifact    & 4,908 & --- \\
\dataset (Section~\ref{sec:scoping})& 647   & 2,752,231 \\
\quad human-authored              &       & 2,473,892 \\
\quad\quad visual PRs             &       & 329,185 \\
\quad\quad non-visual PRs         &       & 2,144,707 \\
\bottomrule
\end{tabular}
\end{table}

\subsection{Preliminary Analysis: Who Embeds Visual Artifacts, and in What Formats?}
\label{sec:preliminary}
We characterize what types of visuals are shared and by whom. We aim to understand the distribution across author types (human developers, bots, and AI agents) and across file formats. We conducted the preliminary analysis on the same 647 repositories we used to create \dataset.

\begin{table}[t]
    \centering
    \caption{Visual PRs and file formats of visual artifacts by author type in \dataset}
    \label{tab:rq0}
    \resizebox{\columnwidth}{!}{%
    \begin{tabular}{@{}l rr rr rr@{}}
        \toprule
        & \multicolumn{2}{c}{\textbf{Human}} & \multicolumn{2}{c}{\textbf{Bot}} & \multicolumn{2}{c}{\textbf{AI agent}} \\
        \cmidrule(lr){2-3} \cmidrule(lr){4-5} \cmidrule(lr){6-7}
        \multicolumn{7}{@{}l}{\textit{Pull requests}} \\
        All PRs                             & \multicolumn{2}{c}{2,473,892} & \multicolumn{2}{c}{270,612}  & \multicolumn{2}{c}{7,727} \\
        Visual PRs                          & \multicolumn{2}{c}{329,185}   & \multicolumn{2}{c}{154,250}  & \multicolumn{2}{c}{1,416}  \\
        \quad \% of the author type's PRs   & \multicolumn{2}{c}{13.3\%}    & \multicolumn{2}{c}{57.0\%}   & \multicolumn{2}{c}{18.3\%} \\
        \quad \% of all visual PRs          & \multicolumn{2}{c}{67.9\%}    & \multicolumn{2}{c}{31.8\%}   & \multicolumn{2}{c}{0.3\%}  \\
        \midrule
        \textit{File formats} & \textbf{Count} & \textbf{\%} & \textbf{Count} & \textbf{\%} & \textbf{Count} & \textbf{\%} \\
        \midrule
        PNG                         & 280,589 & 58.5\%   &   4,458 &  2.7\%   &  8,502 & 41.5\%   \\
        SVG                         & 174,887 & 36.4\%   & 155,496 & 95.7\%   & 11,320 & 55.3\%   \\
        JPG                         &   8,258 &  1.7\%   &   1,338 &  0.8\%   &    376 &  1.8\%   \\
        GIF                         &   6,611 &  1.4\%   &     186 &  0.1\%   &    142 &  0.7\%   \\
        MOV                         &   3,488 &  0.7\%   &       8 & $<$0.1\% &      1 & $<$0.1\% \\
        MP4                         &   3,283 &  0.7\%   &      19 & $<$0.1\% &    107 &  0.5\%   \\
        JPEG                        &   2,301 &  0.5\%   &   1,009 &  0.6\%   &     14 &  0.1\%   \\
        WebP                        &     345 &  0.1\%   &      12 & $<$0.1\% &      9 & $<$0.1\% \\
        WebM                        &     211 & $<$0.1\% &      16 & $<$0.1\% &      4 & $<$0.1\% \\
        TIFF                        &       1 & $<$0.1\% &       0 & ---      &      0 & ---      \\
        \midrule
        \textbf{Total}              & \textbf{479,974} & \textbf{100\%} & \textbf{162,542} & \textbf{100\%} & \textbf{20,475} & \textbf{100\%} \\
        \bottomrule
    \end{tabular}%
    }
\end{table}

\textbf{Format resolution.} Many URLs, particularly those of files uploaded directly to GitHub, do not include a file extension. We therefore determined the format of each visual artifact in two steps: (1)~\emph{Content type}: we requested each URL and mapped the content type reported by the server (e.g., \texttt{image/png} or \texttt{video/mp4}) to the corresponding format. (2)~\emph{File signature}: when the server reported no content type or only a generic one (e.g., \texttt{application/octet-stream}), we downloaded the first 2,048 bytes of the file and compared them with known file signatures, the byte sequences at the start of a file that identify its format. This step also distinguishes formats with similar internal structures, such as MP4 and MOV. Of the  989,807 visual URLs in the corpus, we resolved 662,991 (66.98\%). The remaining URLs could not be resolved because of link rot, expired access tokens on GitHub's Camo image proxy, HTTP errors (e.g., 403 and 404), or connection timeouts, which is expected for media links in PRs spanning up to five years. Table~\ref{tab:rq0} reports visual PRs and file formats by author type. Our main findings are presented below.

\noindent\textbf{Finding 0.1: Bots embed visual artifacts frequently, but almost exclusively as SVG files.} Visual artifacts appear in 57.0\% of bot-authored PRs, compared with 13.3\% of human-authored PRs, and 95.7\% of bot visual artifacts are SVG files, which suggests automatically generated content such as status badges and coverage charts.

\noindent\textbf{Finding 0.2: Most human visual artifacts are PNG files, human SVG files are mostly decorative.} PNG files, a format typical of screenshots, account for 58.5\% of human visual artifacts, and SVG files account for a further 36.4\%.
To check whether human-embedded SVG files resemble bot output, we inspected a random sample of 100 of them: 98\% were badges, logos, icons, or other generated graphics rather than depictions of the change.

\noindent\textbf{Finding 0.3: AI agents author too few visual PRs for separate analysis.} Agents authored  7,727 PRs in the corpus, of which  1,416 are visual PRs (0.3\% of all visual PRs). Their visual artifacts mix SVG and PNG files, suggesting both generated graphics and screenshots.

\noindent\textbf{Finding 0.4: GIFs and videos are rare for every author type.} MP4, MOV, and WebM files together account for at most 1.5\% of any author type's visual artifacts, and GIFs for 1.4\% of human visual artifacts.

\subsection{Sampling from \dataset for answering RQ1-3.} \label{sec:sample_set} Based on the preliminary analysis (Section~\ref{sec:preliminary}), we study human-authored PRs only, which is why \dataset{} is restricted to them. Its 329,185 visual PRs contain 479,974 visual artifacts. We examine three media formats: \emph{images} (PNG, JPG/JPEG, and WebP), \emph{GIFs}, and \emph{videos} (MP4, MOV, and WebM). We exclude SVG files, most of which are badges, logos, or icons rather than depictions of the change (Finding~0.2), and the single TIFF file in the corpus. This leaves 305,086 visual artifacts: 291,493 images, 6,611 GIFs, and 6,982 videos. Because GIFs and videos are rare (Finding~0.4), we sample each media format separately rather than in proportion to its frequency. Following Kuramoto et al.~\cite{kuramoto2024understanding}, we determine each sample size with the formula of Krejcie and Morgan~\cite{krejcie1970determining} at a 95\% confidence level and a 5\% margin of error, and randomly draw 384 images, 364 GIFs, and 365 videos ($N = 1{,}113$). These 1,113 artifacts come from 891 distinct visual PRs. No single repository dominates the sample: the most represented repository accounts for at most 10.9\% of PRs within any media format, the top five repositories together never exceed 32\%, and the median number of sampled PRs per repository is 1 for images and GIFs and 2 for videos. RQ1 and RQ2 analyze the 1,113 artifacts. RQ3 compares the 891 visual PRs with 1,782 non-visual PRs from \dataset, matched as described in Section~\ref{sec:rq3}.


\section{Research Questions and Results}\label{rq}

\subsection{RQ1: \rqone}
\label{sec:rq1}

\noindent\textbf{Motivation.} Section~\ref{sec:preliminary} characterized visual artifacts by author type and file format (Table~\ref{tab:rq0}), but a format does not reveal what a visual artifact shows. A PNG file may show a UI, an error message, or a log, each conveys different information to reviewers. RQ1 therefore examines what developers show in the visual artifacts of their PR descriptions. Answering this requires judging the subject of a visual artifact rather than listing everything visible in it: a screenshot of an IDE may display a large block of code while the developer’s intent is to point to a tooltip or a debugger state. Moreover, GIFs and videos can show change over time, whereas images cannot, so what developers show is likely to differ across media formats. Characterizing this content, and how it varies by format, can help researchers and tool builders support visual-related communication in code review.


\noindent\textbf{Approach.} We analyzed the 1,113 sampled visual artifacts (Section~\ref{sec:sample_set}) and built a taxonomy of content types through two rounds of open coding. In the first round, one author coded 100 visual artifacts randomly drawn from \dataset, using the taxonomy of Kuramoto et al.~\cite{kuramoto2024understanding} as a starting point, and derived an initial codebook. The authors then discussed and refined the codebook. In the second round, two authors independently labeled 90 artifacts randomly drawn from the RQ1 sample following the codebook, reaching a Krippendorff's $\alpha$ of 0.82, above the 0.80 threshold for reliable coding~\cite{krippendorff2019content}. The final taxonomy (Table~\ref{tab:taxonomy}) contains ten content types and an Others category for artifacts that fit none of them and have counts below five (following a common practice~\cite{agresti2007introduction}). The same two authors then divided the remaining 1,023 artifacts between them. To test whether the content type distribution differs across three media formats, we applied a chi-square test.

\begin{table}[t]
    \centering
    \caption{Distribution of content types across media formats.  }
    \label{tab:rq1-results}
    \resizebox{\columnwidth}{!}{%
    \begin{tabular}{@{}l rr rr rr@{}}
        \toprule
        & \multicolumn{2}{c}{\textbf{Image} ($n = 384$)} & \multicolumn{2}{c}{\textbf{GIF} ($n = 364$)} & \multicolumn{2}{c}{\textbf{Video} ($n = 365$)} \\
        \cmidrule(lr){2-3} \cmidrule(lr){4-5} \cmidrule(lr){6-7}
        \textbf{Content type} & \textbf{\#} & \textbf{\%} & \textbf{\#} & \textbf{\%} & \textbf{\#} & \textbf{\%} \\
        \midrule
        Interaction \& State & 142 & 37.0\% & 231 & 63.5\% & 291 & 79.7\% \\
        Visual \& Style      &  66 & 17.2\% &  32 &  8.8\% &  20 &  5.5\% \\
        Rendered Output      &  57 & 14.8\% &  33 &  9.1\% &  14 &  3.8\% \\
        Log                  &  32 &  8.3\% &  19 &  5.2\% &   5 &  1.4\% \\
        Code                 &  21 &  5.5\% &  13 &  3.6\% &  11 &  3.0\% \\
        Documentation        &  20 &  5.2\% &   5 &  1.4\% &   2 &  0.5\% \\
        Error Message        &  17 &  4.4\% &   3 &  0.8\% &   1 &  0.3\% \\
        Layout \& Spatial    &   9 &  2.3\% &   7 &  1.9\% &   6 &  1.6\% \\
        State Change         &   1 &  0.3\% &  13 &  3.6\% &  13 &  3.6\% \\
        Diagram              &   8 &  2.1\% &   3 &  0.8\% &   0 &  0.0\% \\
        Others               &  11 &  2.9\% &   5 &  1.4\% &   2 &  0.5\% \\
        \midrule
        \textbf{Total}       & \textbf{384} & \textbf{100\%} & \textbf{364} & \textbf{100\%} & \textbf{365} & \textbf{100\%} \\
        \bottomrule
    \end{tabular}%
    }
\end{table}


\noindent\textbf{Results.} Table~\ref{tab:rq1-results} presents the distribution of content types across the three media formats. The distribution differs significantly across media formats based on the test result ($\chi^2 = 149.56$, $df = 10$, $p < 0.001$, Cram’er’s $V = 0.259$). Below we summarize the main findings from the observed distributions.

\noindent\textbf{Finding 1.1: User interface (UI) content dominates all three media formats, mainly through \ct{Interaction \& State}.} The three UI content types (\ct{Interaction \& State}, \ct{Visual \& Style}, and \ct{Layout \& Spatial}) together account for 56.5\% of images (217/384), 74.2\% of GIFs (270/364), and 86.8\% of videos (317/365). Most of this share comes from \ct{Interaction \& State}, which shows what happens during or after a user action, such as a click, hover, or drag. It is the most frequent content type in every format, accounting for 37.0\% of images, 63.5\% of GIFs, and 79.7\% of videos. \ct{Visual \& Style} contributes mainly to images (17.2\%), whereas \ct{Layout \& Spatial} remains rare in all formats (at most 2.3\%). The higher shares in GIFs and videos may reflect that motion can capture interactive behavior as it unfolds, whereas an image can capture only the state after an action, such as an opened menu or a displayed tooltip.

\noindent\textbf{Finding 1.2: Beyond UI content, images span a wide range of content types, led by program output.} The distribution of content types differs significantly across media formats ($\chi^2 = 149.56$, $df = 10$, $p < 0.001$, Cram’er’s $V = 0.259$). Non-UI content accounts for 43.5\% of images, compared with 25.8\% of GIFs and 13.2\% of videos. In images, the largest non-UI category is program output (27.9\%, 107/384), comprising \ct{Rendered Output} (14.8\%), \ct{Log} (8.3\%), and \ct{Error Message} (4.4\%), followed by \ct{Code} (5.5\%), \ct{Documentation} (5.2\%), and \ct{Diagram} (2.1\%). In GIFs and videos, program output accounts for only 18.7\% and 9.0\%, respectively, and no non-UI content type exceeds 10\%. The one exception to this pattern is \ct{State Change}, which shows backend behavior triggered through the user interface: it accounts for 3.6\% of GIFs and of videos but only 0.3\% of images, although these counts are small. Images thus serve as a general-purpose medium for many kinds of content, whereas GIFs and videos concentrate on UI behavior (Finding~1.1).



\noindent\textbf{Finding 1.3: About a quarter of images show textual content.} Four content types depict content that is itself text: \ct{Log}, \ct{Code}, \ct{Documentation}, and \ct{Error Message}. Together, they account for 23.4\% of images (90/384), compared with 11.0\% of GIFs (40/364) and 5.2\% of videos (19/365). Text shown in an image cannot be searched, copied, or processed by screen readers and text-based review tools. We did not examine whether this text could have been included directly in the description, but its prevalence indicates that a substantial share of image content is inaccessible to text-based analysis.

\begin{tcolorbox}
\textbf{RQ1 Summary:} 
The content of visual artifacts differs significantly across media formats. UI content dominates all three formats (56.5\% of images, 74.2\% of GIFs, and 86.8\% of videos), mainly through \ct{Interaction \& State}. GIFs and videos concentrate on UI behavior, whereas images also cover a wide range of other content, led by program output; about a quarter of images show textual content such as logs, code, or error messages.
\end{tcolorbox}

\begin{table*}[h]
    \centering
    \caption{Taxonomy of content types and purposes for visual artifacts in PR descriptions, Content types are labeled per visual artifact \& purposes are labeled per PR.}
    \label{tab:taxonomy}
    \scriptsize
    \begin{tabular}{@{}cl p{0.15\textwidth} p{0.72\textwidth}@{}}
        \toprule
        & \textbf{Code} & \textbf{Label} & \textbf{Definition} \\
        \midrule

        \multirow{22}{*}{\rotatebox[origin=c]{90}{\textbf{Content type}}}
        & T1 & Layout \& Spatial
          & The position, size, alignment of UI elements. focus on spatial relationship between elements. \\[3pt]

        & T2 & Visual \& Style
          & The appearance of user interface elements, such as color, font, border, opacity, or theme, rather than their layout. \\[3pt]

        & T3 & Interaction \& State
          & What happens during or after a user action, such as a click, hover, focus, drag, or scroll. \\[3pt]

        & T4 & Diagram
          & A flowchart, architecture, or other diagram where the structure itself is the content. \\[3pt]

        & T5 & Rendered Output
          & Output that a program produced when executed, such as a data visualization, a generated document, or creative output. \\[3pt]

        & T6 & Error Message
          & A single failure message produced by a program (but not a sequence). \\[3pt]

        & T7 & Log
          & Multiple timestamped or ordered lines recording what a program did while running. \\[3pt]

        & T8 & State Change
          & A backend change triggered through the UI, such as a database update shown after a button click. \\[3pt]

        & T9 & Code
          & Source code, where the code itself is the subject. \\[3pt]

        & T10 & Documentation
          & Rendered developer documentation, such as API references, READMEs, or component property tables. \\[3pt]

        & -- & Others
          & Visual content that fits none of the above content types. \\

        \midrule
        \midrule

        & \textbf{Code} & \textbf{Label} & \textbf{Definition} \\
        \midrule

        \multirow{12}{*}{\rotatebox[origin=c]{90}{\textbf{Purpose}}}
        & P1 & Introduce
          & The PR adds something that did not exist before. No prior state is needed to understand the visual artifacts. \\[3pt]

        & P2 & Improve
          & The PR improves something that was not broken, and the visual artifact shows the better new state. \\[3pt]

        & P3 & Verify
          & The PR documents that something is broken or behaves unexpectedly. The visual artifacts serve as evidence of the correction. \\[3pt]

        & P4 & Prior State -- Bug
          & The visual artifacts show the broken or unexpected state before the fix, establishing the context for the change. \\[3pt]

        & P5 & Prior State -- No Bug
          & Shows what was there before the change, not what was wrong, as a baseline for comparison. \\[3pt]

        & -- & Others
          & A purpose that fits none of the above purposes. \\

        \bottomrule
    \end{tabular}
\end{table*}

\subsection{RQ2:\rqtwo}
\label{sec:rq2}

\noindent\textbf{Motivation.} RQ1 characterized what developers show in the visual artifacts of their PR descriptions, but the same content can serve different purposes. A screenshot of a dropdown menu, for example, may present a newly added option, confirm that a previously misaligned menu now renders correctly, or show the existing menu before a redesign. Each purpose calls for a different response from reviewers: a new feature must be evaluated, a fix must be confirmed, and a prior state provides context for judging the change. RQ2 therefore examines \emph{why} developers include visual artifacts in their PR descriptions. Understanding these purposes clarifies the role visual artifacts play in PR communication.


\noindent\textbf{Approach.} We used the same 1,113 visual artifacts sample set (Section~\ref{sec:sample_set}) and labeled the purpose of each one, i.e., why the author included it in the PR description. We inferred the purpose from the PR description rather than the visual artifact alone. The two authors labeled purposes and content types at the same time, following the two-round procedure described for RQ1 (Section~\ref{sec:rq1}). In the second round, inter-rater agreement on purposes reached a Krippendorff’s $\alpha$ of 0.902. The final taxonomy contains six purposes (Table~\ref{tab:taxonomy}). As in RQ1, we cross-tabulated the labeled artifacts by media format and purpose and applied a chi-square test of independence~\cite{pearson1900}, reporting Cram’{e}r’s $V$ as the effect size~\cite{cohen1988statistical}. 

\begin{table}[t]
    \centering
    \caption{Distribution of purposes across media formats. }
    \label{tab:rq2-results}
    \resizebox{\columnwidth}{!}{%
    \begin{tabular}{@{}l rr rr rr@{}}
        \toprule
        & \multicolumn{2}{c}{\textbf{Image} ($n = 384$)} & \multicolumn{2}{c}{\textbf{GIF} ($n = 364$)} & \multicolumn{2}{c}{\textbf{Video} ($n = 365$)} \\
        \cmidrule(lr){2-3} \cmidrule(lr){4-5} \cmidrule(lr){6-7}
        \textbf{Purpose} & \textbf{\#} & \textbf{\%} & \textbf{\#} & \textbf{\%} & \textbf{\#} & \textbf{\%} \\
        \midrule
        Introduce             & 141 & 36.7\% & 169 & 46.4\% & 180 & 49.3\% \\
        Verify                & 150 & 39.1\% & 138 & 37.9\% & 156 & 42.7\% \\
        Prior State -- Bug    &  24 &  6.3\% &  21 &  5.8\% &  15 &  4.1\% \\
        Prior State -- No Bug &  37 &  9.6\% &  27 &  7.4\% &  10 &  2.7\% \\
        Improve               &  19 &  4.9\% &   5 &  1.4\% &   3 &  0.8\% \\
        Others                &  13 &  3.4\% &   4 &  1.1\% &   1 &  0.3\% \\
        \midrule
        \textbf{Total}        & \textbf{384} & \textbf{100\%} & \textbf{364} & \textbf{100\%} & \textbf{365} & \textbf{100\%} \\
        \bottomrule
    \end{tabular}%
    }
\end{table}

\noindent\textbf{Results.} Table~\ref{tab:rq2-results} presents the distribution of purposes across the three media formats. The distribution differs significantly across media formats, although the effect is small ($\chi^2 = 51.92$, $df = 10$, $p < 0.001$, Cram\'{e}r's $V = 0.153$).

\noindent\textbf{Finding 2.1: Most visual artifacts show the outcome of the change, either demonstrating a functionality that didn't exist before (\pp{Introduce}) or showing evidence that the change works (\pp{Verify}).} \pp{Introduce} and \pp{Verify} together account for 75.8\% of images (291/384), 84.3\% of GIFs (307/364), and 92.1\% of videos (336/365). Including \pp{Improve}, the purposes that present the state after the change account for 80.7\%, 85.7\%, and 92.9\%, respectively, whereas artifacts introduces a problem (\pp{Prior State -- Bug}) account for only 4.1--6.3\%. This reverses the role that visual artifacts play in issue reports, where they mainly show program behavior and the user interface to introduce a problem~\cite{kuramoto2024understanding}: in PR descriptions, they mainly present the solution. We discuss what this means for reviewers in Section~\ref{dicussion}.

\noindent\textbf{Finding 2.2: Purposes vary less across media formats than content types do; the main shift is a rising share of \pp{Introduce}.} The distribution of purposes differs significantly across media formats, but with a small effect ($\chi^2 = 51.92$, $df = 10$, $p < 0.001$, Cram’er’s $V = 0.153$), smaller than for content types in RQ1 ($V = 0.259$). The share of \pp{Introduce} rises from 36.7\% of images to 46.4\% of GIFs and 49.3\% of videos, whereas \pp{Verify} stays between 37.9\% and 42.7\%. As a result, \pp{Verify} is slightly more frequent than \pp{Introduce} in images, whereas \pp{Introduce} is the most frequent purpose in GIFs and videos. GIFs and videos thus more often accompany new functionality, such as a new interaction flow, which motion can demonstrate.

\noindent\textbf{Finding 2.3: Showing the prior state is uncommon, especially in videos.} The two \pp{Prior State} purposes together account for 15.9\% of images (61/384), 13.2\% of GIFs (48/364), and 6.8\% of videos (25/365). \pp{Prior State -- No Bug} declines most, from 9.6\% of images to 2.7\% of videos, while \pp{Prior State -- Bug} declines from 6.3\% to 4.1\%. Visual artifacts therefore rarely give reviewers a visual baseline for the change; when they do, it is mostly in images.

\begin{tcolorbox}
\textbf{RQ2 Summary:} 
The purposes of visual artifacts differ across media formats, but less than their content does. Most visual artifacts show the outcome of the change, mainly to \pp{Introduce} something new or \pp{Verify} that the change works (75.8\% of images, 84.3\% of GIFs, and 92.1\% of videos). \pp{Introduce} becomes more frequent from images to videos, and showing the prior state is uncommon, especially in videos (6.8\%).
\end{tcolorbox}

\subsection{RQ3: \rqthree}\label{sec:rq3}

\noindent\textbf{Motivation.} 
Developers invest effort in producing screenshots and recordings, and many projects request them in PR descriptions templates\footnote{https://github.com/kodadot/nft-gallery/pull/4374}. Yet little is known about whether the review process itself differs when a PR contains visual artifacts. If visuals attract additional reviewer attention, we would expect to see more comments, more reviewers, or longer post-response iteration compared with similar PRs that lack visuals. If, on the other hand, visuals merely substitute for textual explanation, review activity should be comparable once change size is controlled for. RQ3 therefore examines whether review characteristics differ between PRs with and without visual artifacts in their descriptions, and if so, along which dimensions from efficiency, timing, feedback volume, or reviewer participation differ.


\begin{table*}[t]
\centering

\caption{Codebook of review discussion codes and their definitions.}
\label{tab:rq3-codebook}
\scriptsize
\begin{tabular}{@{}c l l p{0.70\textwidth}@{}}

\toprule
& \textbf{Code} & \textbf{Label} & \textbf{Definition} \\
\midrule
\multirow{7}{*}{\rotatebox[origin=c]{90}{\textbf{Comment Topic}}}
& CL & Code-logic       & About algorithms, control flow, error handling, edge cases, performance, or security. \\
& CS & Code-style       & About naming, formatting, conventions, linting, or import order. \\
& VO & Visual-output    & About appearance of the rendered result: spacing, alignment, colors, fonts, layout, or size. \\
& VB & Visual-behavior  & About how the UI behaves: interactions, animations, transitions, responsiveness, or state changes. \\
& PR & Process          & About the PR process: CI, tests, merge conflicts, or requesting more context. \\
& AP & Approval         & LGTM, looks good, approved.  No substantive feedback. \\
& CQ & Clarification    & Asking for more information without making a judgment. \\
\midrule
\multirow{3}{*}{\rotatebox[origin=c]{90}{\textbf{Vis.\ Ref.}}}
& E & Explicit          & Names the visual artifact directly (e.g., ``In the screenshot\ldots'', ``The GIF shows\ldots''). \\
& I & Implicit          & Describes something the commenter could only know from looking at the visual, without naming it. \\
& N & None              & Comment is entirely about code, process, or approval with no visual reference. \\
\midrule
\multirow{5}{*}{\rotatebox[origin=c]{90}{\textbf{Entry Point}}}
& VF      & Visual-first  & First substantive comment references the visual; code comments follow later. \\
& CF      & Code-first    & First comment is inline on code; visual comments follow later. \\
& VO-only & Visual-only   & All comments are about the visual or top-level. No inline code engagement. \\
& CO-only & Code-only     & All comments are about code or process. No visual reference anywhere. \\
& SM      & Simultaneous  & First review submission contains both inline code and visual-referencing comments. \\
\midrule
\multirow{6}{*}{\rotatebox[origin=c]{90}{\textbf{Revision Type}}}
& CLF & Code-logic-fix        & Changed functionality or fixed a bug the reviewer identified. \\
& CSF & Code-style-fix        & Renamed, reformatted, or adjusted convention in response to a style comment. \\
& VOF & Visual-output-fix     & Changed CSS, layout, or template to alter the visual output. \\
& VAU & Visual-artifact-update & Author posted a new screenshot or GIF showing revised output. \\
& TF  & Test/CI-fix           & Added test, fixed build, or addressed CI failure. \\
& NR  & No-revision           & No commits after first review. \\
& UN & Undetermined & Revision type cannot be determined from the commit messages. \\
\bottomrule
\end{tabular}
\end{table*}

\noindent\textbf{Approach.} To answer RQ3, we perform the following steps.

\noindent\textit{Step 1: Form control and treatment groups.} Because review outcomes are observed per PR rather than per artifact, we conduct RQ3 analysis at the PR level. The treatment group consists of the 891 distinct visual PRs that contain the 1,113 sampled artifacts (Section~\ref{sec:sample_set}). To isolate the effect of visual artifacts from other PR characteristics, we pair each visual PR with non-visual PRs (i.e., PRs whose descriptions contain no visual artifact) that resemble it, following standard matching practice for observational data. We match exactly on two categorical attributes. First, we match on \emph{repository}, which controls for project-specific review norms, team size, and contribution volume. Second, we match on \emph{PR type}, the purpose of the change (e.g., \texttt{Feature}, \texttt{Fix}, \texttt{Refactor}), as assigned by the classifier of Zhong et~al.~\cite{zhong2026human}, because review effort differs systematically across change purposes. Within each exact-match stratum, we select the two nearest non-visual PRs by change size, measured as the Euclidean distance over $\log(\textit{churn})$ and $\log(\textit{files changed})$. We use these two covariates because they are the primary inputs to the code-review efficiency measure of Zhong et~al.~\cite{zhong2026human}, and we log-transform them because both are heavily right-skewed. Distance-based matching avoids the arbitrary bin boundaries that exact matching on continuous covariates would require. We use a 1:2 ratio because adding a second control reduces variance with little increase in bias, whereas returns diminish beyond that~\cite{austin2010ratio}. This procedure yields 1,782 control PRs, and together with the 891 treatment PRs, our RQ3 dataset comprises 2,673 PRs.

\noindent\textit{Step 2: Measuring review characteristics.} For each of the 2,673 matched PRs (891 visual and 1,782 control), we computed ten metrics in five dimensions (Table~\ref{tab:rq3-factors}):
\begin{enumerate}[label=(\arabic*), leftmargin=*, nosep] 
  \item \emph{Standard efficiency}: hours to merge and review efficiency, defined as days to merge per thousand lines of churn~\cite{zhong2026human}.
  \item \emph{Review latency}: time to first response (TTFR), the hours from PR creation to the first comment or review by someone other than the author, and post-response time, the hours from the first response to merge. 
  \item \emph{Review iteration}: the number of reviews submitted on a PR returned by GitHub's Pull Requests Reviews API, each with a state of \texttt{APPROVED}, \texttt{CHANGES\_REQUESTED}, or \texttt{COMMENTED}. We exclude reviews by the PR author. 
  \item \emph{Feedback volume}: the number of inline comments on code, pull request comments, and review comments, and their total.
  \item \emph{Reviewer participation}: the number of distinct reviewers and the number of comments per reviewer.
\end{enumerate}


\noindent\textit{Step 3: Categorizing review discussions.} The metrics in Step~2 capture quantitative aspects of review, but not qualitative aspects, such as what the discussion addresses. To categorize review discussions, one author developed a taxonomy by manually examining both issue, review, and inline comments from 30 randomly sampled PRs from both treatment and control groups. The authors then discussed and confirmed the taxonomy. The final taxonomy categorizes review discussions along four dimensions (Table~\ref{tab:rq3-codebook}): the \emph{topic} of each comment (7 codes); whether a comment \emph{references the visual artifact} explicitly, implicitly, or not at all (3 codes); the \emph{entry point} of the review, i.e., whether the first substantive comment addresses the code, the visual artifact, or both (5 codes); and the \emph{revision} that followed the review (6 codes). Because manual labeling of the 10,031 comments, including issue, inline, and review comments, from 2,673 PRs was infeasible, we used an LLM (GPT-4.1-mini) as a classifier to further scale the analysis, prompting it with the codebook definitions in Table~\ref{tab:rq3-codebook}, as the definition detail is in the prompt. It assigned each comment a topic and a visual-reference, and each PR an entry point and a revision type. To assess its reliability, one author independently labeled a stratified sample of 40 PRs (30 visual and 10 control), selected to cover all visual content types. Per-dimension agreement was: comment topic 91.4\% (Cohen's $\kappa = 0.88$), visual reference 100\%, entry point 100\% ($\kappa = 1.00$), and revision type 92.5\% ($\kappa = 0.86$). Given this level of agreement, we used the LLM labels for the full sample.

\noindent\textit{Step 4: Statistical comparison.} We compared visual and paired non visual PRs on each measurement mentioned in Step~2 and Step~3 using the Mann–Whitney U test and report Cliff’s $\delta$ as the effect size \cite{mann1947test}. 

\noindent\textbf{Results.} \noindent\textbf{Finding 3.1: Time to merge does not differ significantly between visual and non-visual PRs, but visual PRs take longer after the first response.}
Hours to merge (median 18.36 vs.\ 15.22~h, $p = 0.111$) and review efficiency (median 105.17 vs.\ 94.37~h/KLOC, $p = 0.311$) do not differ significantly between the two groups (Table~\ref{tab:rq3-factors}). Among PRs that received at least one response, time to first response does not differ significantly either (median 3.15 vs.\ 2.92~h, $p = 0.483$). The difference appears after the first response: post-response time is 72\% longer for visual PRs (median 5.90 vs.\ 3.43~h, $p = 0.042$), although the effect size is negligible ($\delta = 0.053$; we treat $|\delta| < 0.147$ as negligible~\cite{romano2006appropriate}).

\begin{table}[!t]
\centering
\caption{Review characteristics of visual PRs and their matched control PRs. }
\label{tab:rq3-factors}
\resizebox{\columnwidth}{!}{%
\begin{tabular}{l rr rrr}
\toprule
 & \textbf{Visual} & \textbf{Control} & & & \\
\textbf{Metric} & \textbf{Value} & \textbf{Value}
  & \textbf{$\Delta$\%} & \textbf{$p$} & \textbf{Cliff's $\delta$} \\
\midrule
\multicolumn{6}{l}{\textit{Standard metrics (median)}} \\
Review efficiency (h/KLOC) & 105.17 & 94.37 & +11\%  & 0.311  & 0.024 \\
Hours-to-merge (h)         &  18.36 & 15.22 & +21\%  & 0.111  & 0.038 \\
\midrule
\multicolumn{6}{l}{\textit{Timing (median)}} \\
TTFR (h)                          &  3.15 &  2.92 &  +8\% & 0.483  & 0.018 \\
\textbf{Post-response (h)}        &  \textbf{5.90} &  \textbf{3.43}
  & \textbf{+72\%} & \textbf{0.042} & \textbf{0.053} \\
\midrule
\multicolumn{6}{l}{\textit{Iteration (mean)}} \\
\textbf{Review rounds}     &  \textbf{1.30} &  \textbf{1.12}
  & \textbf{+16\%} & \textbf{$<$0.0001} & \textbf{0.091} \\
\midrule
\multicolumn{6}{l}{\textit{Feedback volume (mean)}} \\
\textbf{Total comments}    & \textbf{5.89} & \textbf{4.13}
  & \textbf{+43\%} & \textbf{$<$0.0001} & \textbf{0.134} \\
\textbf{Inline comments}   & \textbf{2.35} & \textbf{1.48}
  & \textbf{+59\%} & \textbf{$<$0.0001} & \textbf{0.087} \\
\textbf{Top-level comments} & \textbf{3.54} & \textbf{2.66}
  & \textbf{+33\%} & \textbf{$<$0.0001} & \textbf{0.133} \\
\midrule
\multicolumn{6}{l}{\textit{Reviewer participation (mean)}} \\
\textbf{Reviewer count}     & \textbf{1.58} & \textbf{1.35}
  & \textbf{+17\%} & \textbf{$<$0.0001} & \textbf{0.105} \\
\textbf{Comments per reviewer} & \textbf{3.41} & \textbf{2.81}
  & \textbf{+21\%} & \textbf{$<$0.0001} & \textbf{0.116} \\
\bottomrule
\end{tabular}%
}
\end{table}

\noindent\textbf{Finding 3.2: Visual PRs receive more comments, more reviewers, and more review rounds than non-visual PRs, although the effect sizes are negligible.}
All six iteration, feedback, and participation metrics differ significantly ($p < 0.0001$), but all effect sizes are negligible ($\delta \le 0.134$; Table~\ref{tab:rq3-factors}). Visual PRs receive 43\% more comments in total (mean 5.89 vs.\ 4.13), and both inline comments (mean 2.35 vs.\ 1.48, +59\%) and top-level comments (mean 3.54 vs.\ 2.66, +33\%) are higher. Visual PRs also have 17\% more reviewers (mean 1.58 vs.\ 1.35), and each reviewer writes more comments (mean 3.41 vs.\ 2.81, +21\%). Review rounds have the same median in both groups (1 vs.\ 1), but the mean is 16\% higher for visual PRs (1.30 vs.\ 1.12).

\begin{table}[t]
\centering
\caption{ Coding summary at Comment-level using LLM labeling.}
\label{tab:rq3-qualitative}
\scriptsize
\setlength{\tabcolsep}{4pt}
\begin{tabular}{@{}lrr@{\hspace{12pt}}lrr@{}}
\toprule
\multicolumn{3}{c}{\textbf{Per comment}} & \multicolumn{3}{c}{\textbf{Per PR}} \\
\cmidrule(r){1-3} \cmidrule(l){4-6}
Code & Visual & Control & Code & Visual & Control \\
$n$  & 4{,}200 & 5{,}831 & $n$ & 891 & 1{,}782 \\
\midrule
\multicolumn{3}{@{}l}{\textit{Comment topic}} & \multicolumn{3}{l@{}}{\textit{Entry point}} \\
CL & \textbf{30.3} & \textbf{29.3} & CO-only & \textbf{59.7} & \textbf{61.4} \\
AP & 27.8 & 32.6 & CF      & \textbf{6.1} & \textbf{2.2} \\
PR & 13.1 & 15.4 & VO-only & \textbf{4.8} & 1.0 \\
CS & \textbf{11.8} & \textbf{9.9} & VF & 3.9 & 0.7 \\
CQ & 10.5 &  9.4 & SM      & 1.1 & 0.5 \\
VO & \textbf{4.3} & \textbf{2.2} & \multicolumn{3}{l@{}}{\textit{Revision type}} \\
VB & \textbf{2.1} & \textbf{1.1} & NR & \textbf{60.8} & \textbf{61.2} \\
\multicolumn{3}{@{}l}{\textit{Visual reference}} & CLF & \textbf{19.3} & \textbf{24.0} \\
N  & \textbf{92.2} & \textbf{94.8} & CSF & 8.2 & 7.3 \\
E  & \textbf{6.0} & \textbf{4.3} & VOF & \textbf{4.7} & \textbf{1.2} \\
I  & \textbf{1.9} & \textbf{0.9} & TF  & 2.4 & 2.7 \\
   &      &      & UN      & 4.3 & 3.4 \\
\bottomrule
\end{tabular}
\end{table}

\noindent\textbf{Finding 3.3: Only a small share of review comments reference the visual artifact, and review typically starts from the code.} 7.9\% of comments on visual PRs refer to the visual, either explicitly (6.0\%) or implicitly (1.9\%). 92.2\% make no reference to it. 65.8\% of visual PRs receive a code-first comment (59.7\% code only, 6.1\% code then visual), close to the 63.6\% for controls, and only 4.8\% begin with a comment on the visual alone.

\noindent\textbf{Finding 3.4: Code remains the main review topic, but visual-specific comments are about twice as common on visual PRs.} Code logic and style account for 42.1\% of comments on visual PRs and 39.3\% on controls. Comments on visual output and visual behaviour make up 6.4\% of comments on visual PRs (4.3\% and 2.1\%) versus 3.3\% on controls (2.2\% and 1.1\%). 

\noindent\textbf{Finding 3.5: Visual PRs are revised as often as non-visual PRs, but visual-output fixes are about four times as common.} The share of PRs revised after the first review is nearly identical (39.2\% vs.\ 38.8\%), and code-logic fixes are the most frequent revision in both groups (19.3\% vs.\ 24.0\%). Visual-output fixes, however, occur in 4.7\% of visual PRs versus 1.2\% of controls.
 
\begin{tcolorbox}
\textbf{Summary for RQ3:} Visual PRs do not differ significantly from matched non-visual PRs in time to merge or review efficiency. They receive 43\% more comments and 17\% more reviewers, and take longer from first response to merge, but all effect sizes are negligible. Although code dominates review in both groups, comments on visual PRs refer to the visual about 1.5 times as often as comments on controls (7.8\% vs.\ 5.2\%), and visual-specific topics are twice as common (6.4\% vs.\ 3.3\%).
\end{tcolorbox}

\section{Discussion}\label{dicussion}


\noindent\textbf{Implication 1: PR templates and hosting platforms should capture the state before the change, not only the state after it.} Taxonomies of issue images record what an image shows or what the reporter wants~\cite{nayebi2020eye,zhan2026mmissueloc}, and ours also records whether it shows the state before or after the change. In PR descriptions, 80.7\%--92.9\% of visual artifacts show the state after the change and 6.8\%--15.9\% the state before (Findings~2.1 and~2.3). Yet \pp{Verify}, the most frequent purpose of images (39.1\%), claims that the change works, and \pp{Improve} claims that the new state is better. Neither claim can be checked without the earlier state. A screenshot of a correctly rendered dropdown menu does not show that it was misaligned before, so the reviewer must run the old code or trust the author. Pirouzkhah et al.\ list screenshots as optional in PR descriptions of user interface changes~\cite{pirouzkhah2026value}, and our results show what they should contain. PR templates should ask for a screenshot of the same view before and after any change to existing behavior or appearance. Hosting platforms can produce the earlier state for every project. GitHub already compares changed image files in three views, including side by side~\cite{github_attaching_files}, but not screenshots of the running application. When a PR changes interface files such as style sheets, the platform can capture and compare the same views before and after the change, as Percy and Chromatic do for projects that set them up.\footnote{\url{https://percy.io}, \url{https://www.chromatic.com}} When a PR closes an issue with a visual artifact, the platform can show it as the earlier state.

\noindent\textbf{Implication 2: Maintainers should ask for visual artifacts to help reviewers catch visual problems, not to make review faster}. Visual issue reports receive faster responses than other issues~\cite{kuramoto2024understanding,nayebi2024image}, but visual PRs do not. Their times to first response and to merge do not differ significantly from matched non-visual PRs (Finding~3.1), although they receive 43\% more comments and 17\% more reviewers (Finding~3.2). What differs is what review addresses. Comments on visual output and behavior make up 6.4\% of comments on visual PRs versus 3.3\% on non-visual ones (Finding~3.5), and fixes to visual output occur in 4.7\% versus 1.2\% of PRs (Finding~3.7). Earlier studies found that 75\% of code review defects do not affect what users see~\cite{mantyla2009types} and 75\% of problems fixed in open-source review concern maintainability rather than functionality~\cite{beller2014modern}. Visual artifacts are thus associated with review of what users see. A screenshot requirement in a PR template is unlikely to shorten review, but our results support it for catching visual problems before merge. Studies that measure only review speed~\cite{zhong2026human} miss this difference and should also count which kinds of problems review fixes.


\noindent\textbf{Implication 3: Logs, code, and error messages in images should also be available as text, pasted by authors or read by multimodal tools.} Like taxonomies of issue images~\cite{nayebi2020eye,zhan2026mmissueloc}, ours includes images of text, which make up 23.4\% of images in PR descriptions (Finding~1.3), namely \ct{Log} (8.3\%), \ct{Code} (5.5\%), \ct{Documentation} (5.2\%), and \ct{Error Message} (4.4\%). Search, screen readers~\cite{wcag21}, and text-only tools and studies of PR descriptions cannot read this text. A developer who later meets the same error cannot find the PR, and tools lose error messages and logs that help locate the code that must change~\cite{zhan2026mmissueloc}. PR templates can ask authors to paste logs, error messages, and code in code blocks and keep images for what text cannot show, which costs little because authors already have the text when taking the screenshot. For the remaining screenshots, tools that read PR descriptions should be multimodal. They can extract the text, as psc2code does for code in programming screen recordings~\cite{bao2020psc2code}, or pass images to multimodal models, and hosting platforms can add the extracted text to search.

\noindent\textbf{Implication 4: AI agents should capture the application before and after their change when they open a PR, and read visual artifacts when they review one.} AI agents add visual artifacts to 18.3\% of their PRs, but 55.3\% are SVG files (Finding~0.3), so whether they show the change is unknown. A human author who attaches a screenshot has seen the change run, but in an agent's PR the reviewer may be the first to see it run, so apart from tests the visual artifact is the reviewer's evidence of how it looks and behaves. Agents that operate a browser or emulator can capture the earlier state that humans leave out (Implication~1) and record how the interface responds to user actions (\ct{Interaction \& State}), as 63.5\% of GIFs and 79.7\% of videos do (Finding~1.1). A review agent that reads only code and text cannot write the 7.8\% of comments that refer to a visual artifact (Finding~3.4) or read text in images (Implication~3). Review agents should be multimodal, receive visual artifacts with the code changes, and be evaluated on such comments.

\noindent\textbf{Implication 5: Benchmark builders should use the visual artifacts of the PR that resolved a task as its expected result.} Current benchmarks take images from issue reports, which show the problem. SWE-bench Multimodal requires an image in the problem statement or tests of each of its 617 tasks~\cite{swebench_multimodal}, and MM-IssueLoc pairs 652 issues with the PRs that resolved them~\cite{zhan2026mmissueloc}. The resolving PR already shows the result (Implication~1), and a benchmark can compare a screenshot of the agent's result with it, as Design2Code compares generated webpages with the originals~\cite{si2025design2code}, to check what a user sees besides running tests. GIFs and videos that \pp{Introduce} new functionality (46.4\% and 49.3\%, Findings~2.2 and~2.3) can also supply feature tasks, whereas SWE-bench Multimodal contains bug-fixing tasks. Builders should exclude bot PRs, which would make up 31.8\% of PRs selected only for having an image in the description, and 95.7\% of whose visual artifacts are SVG files (Finding~0.1). They should skip the 4.7\% of PRs with visual-output fixes after review (Finding~3.7), whose visual artifact may predate these fixes, and store the media files, because 33.0\% of visual URLs could no longer be retrieved (Section~\ref{sec:preliminary}).

\noindent\textbf{Implication 6: Research on PR descriptions should analyze visual artifacts as part of the description, using our dataset and taxonomy.} Studies of PR descriptions analyze their text, and Pirouzkhah et al.\ excluded screenshots because they depend on each PR's context~\cite{pirouzkhah2026value}. Yet visual artifacts appear in 4,908 of the 5,012 repositories we collected (97.9\%) and in 13.3\% of human-authored PRs in our 647 repositories (Table~\ref{tab:dataset}), and our taxonomy describes their content type and purpose, which a text-only analysis misses. Besides excluding bot PRs and storing media files (Implication~5), studies must determine each artifact's format from the file, because URLs of files uploaded to GitHub can lack an extension, and must sample each media format separately. GIFs make up 2.2\% and videos 2.3\% of human visual artifacts in the RQ1 media formats, so a proportional sample of 384 would contain about 8 GIFs and 9 videos, too few to describe content that differs significantly across formats (Finding~1.2).

\section{Threats to Validity} \label{threats}


Our regex-based visual artifact detection may miss visuals embedded via GitHub-specific Markdown extensions or PR comments, and may capture decorative non-communicative visuals such as badges. files from RQ1--RQ3. Link rot across our five-year dataset may cause undercounting of certain media types, but is unlikely to introduce systematic bias, as link decay is independent of visual content type. UI changes that independently attract more reviewer attention, and authors who attach visuals may be more experienced. Our observational design does not support causal claims, and the observed associations may reflect a common cause rather than a direct effect of the visual artifact. In addition, our LLM-assisted coding may introduce systematic biases, but agreement with manual labels on 40 PRs was high (comment topic $\kappa = 0.88$, entry point $\kappa = 1.00$, revision type $\kappa = 0.86$, visual reference 100\%).  Our findings are limited to GitHub, and other platforms differ in media support and review workflows. Our corpus is restricted to mature, actively maintained open-source repositories with established visual communication norms that serve English-speaking communities. This limits the generalizability of our findings to smaller teams, corporate codebases, and non-English communities. Our results reflect a snapshot of practices between August 2021 and August 2026.



\section{Related Work}\label{related}
\noindent \textbf{Pull Request Descriptions and Review Dynamics.} Prior work has examined how the textual content of PR descriptions affects review outcomes. Ram et al.~\cite{ram2018makes} found that a clear change description is one of the strongest predictors of code change reviewability. Pirouzkhah et al.~\cite{pirouzkhah2026value} analyzed 80K PRs across 156 projects and found that stating the type of feedback expected is the single best predictor of PR acceptance and reviewer engagement. their taxonomy of PR description elements identified screenshots as a recommended element but excluded them from regression analysis due to their context-dependency on UI-related changes. Liu et al.~\cite{liu2019automatic} found that over 34\% of PRs in a 333K-PR corpus have empty descriptions and proposed an approach to generate them automatically. Our work complements these studies by examining the \emph{visual} component of PR descriptions, which all prior work either excluded or unanalyzed. Review dynamics more broadly have been studied through large-scale repository mining. Gousios et al.~\cite{gousios2014exploratory} established foundational models of PR latency and acceptance, identifying PR size and submitter reputation as key predictors establishing a baseline against which we compare the topic distribution of comments on visual versus non-visual PRs.

\noindent \textbf{Visual Artifacts in Software Engineering Communication.} The closest line of research to our work studies visual content in issue reports rather than PRs. Kuramoto et al.~\cite{kuramoto2024understanding} studied 34 GitHub repositories and found that issue reports containing images or videos are more likely to describe bugs (44.6\% for images, 71.1\% for videos, vs.\ 32.2\% for non-visual reports), receive their first comment faster, and conclude in less time, their taxonomy of visual issue content serves as a starting point for our own content taxonomy. Tan et al.~\cite{imageR2025} labeled 6,235 Bugzilla issue reports by image type and built a classifier that predicts when an image is needed (F1 = 0.76), finding that 22.5\% of attached images fail to aid resolution. These findings motivate the study of visual content in PRs but do not transfer directly. In an issue report, a visual artifact introduces a problem not yet fixed, whereas in a PR description it accompanies a completed change that reviewers must evaluate.

\noindent \textbf{Taxonomy and Classification of Visual Content.} Several efforts have proposed taxonomies of visual content in software engineering. Kuramoto et al.~\cite{kuramoto2024understanding} defined categories for visual issue reports focusing on program behavior and UI states. Tan et al.~\cite{imageR2025} identified ten image categories in bug reports, including UI screenshots, error logs, and workflow diagrams. Zhan et al.~\cite{zhan2026mmissueloc} refined this into a seven-category scheme that distinguishes UI, error, rendering, diagram, and log content, and introduces relevance levels for issue localization. 
Our work builds on and extends these taxonomies to the PR context, where developers may also show prior states, code diffs, or architectural diagrams in addition to UI behavior, and where the communicative purpose differs from bug reporting.

\section{Conclusion and Future Work}\label{conclusion}
We studied visual artifacts in 329,185 human-authored visual PRs from 647 repositories. What developers show depends on the media format: \textit{Interaction \& State} dominates GIFs (63.5\%) and videos (79.7\%), whereas images also show logs, code, and error messages (23.4\%). Most visual artifacts show the outcome of the change, either new functionality (\textit{Introduce}) or evidence that it works (\textit{Verify}), and few show the state before it. Compared with matched non-visual PRs, visual PRs receive 43\% more review comments and 17\% more reviewers and take 72\% longer from first response to merge, yet time to merge and churn-based review efficiency do not differ. Review centers on code in both groups, but comments on visual PRs address what users see twice as often, and visual-output fixes are four times as common. Visual artifacts are thus associated with closer scrutiny of what users see rather than faster review. Our observational design cannot establish causality. Future work should run controlled experiments to isolate the effect of visual artifacts on review, build classifiers for content type and purpose, and study how reviewers and AI agents use visual artifacts that show the state before a change.




\bibliographystyle{ACM-Reference-Format}
\bibliography{main}
\end{document}